\documentclass[superscriptaddress,aps,pra,twocolumn,preprintnumbers,amsmath,amssymb,floatfix,showpacs]{revtex4-2}

\usepackage{graphicx}% Include figure files
\usepackage{dcolumn}% Align table columns on decimal point
\usepackage{bm}% bold math
\usepackage{amssymb}
\usepackage{subfigure}
\usepackage{xcolor}
\usepackage{stmaryrd}
\usepackage[all]{xy}
\usepackage{textcomp}
\usepackage{xr}

\newcommand{\ket}[1]{\ensuremath{\left|#1\right\rangle}}

\begin{document}

\title{A multiplexed synthesizer for non-Gaussian photonic quantum state generation}

\author{M. F. Melalkia}
\affiliation{Universit\'e C\^ote d'Azur, CNRS, Institut de Physique de Nice (INPHYNI), UMR 7010, Parc Valrose, Nice Cedex 2, France}
\author{J. Huynh}
\affiliation{Universit\'e C\^ote d'Azur, CNRS, Institut de Physique de Nice (INPHYNI), UMR 7010, Parc Valrose, Nice Cedex 2, France}
\author{S. Tanzilli}
\affiliation{Universit\'e C\^ote d'Azur, CNRS, Institut de Physique de Nice (INPHYNI), UMR 7010, Parc Valrose, Nice Cedex 2, France}
\author{V. D'Auria}
\affiliation{Universit\'e C\^ote d'Azur, CNRS, Institut de Physique de Nice (INPHYNI), UMR 7010, Parc Valrose, Nice Cedex 2, France}
\affiliation{Institut Universitaire de France (IUF), France}
\author{J. Etesse}
\affiliation{Universit\'e C\^ote d'Azur, CNRS, Institut de Physique de Nice (INPHYNI), UMR 7010, Parc Valrose, Nice Cedex 2, France}

\date{\today}

%The generation of non-Gaussian photonic states is a challenging task for which complex resources and protocols are usually required. Here, conversely, a large part of the protocol is performed in parallel, by using a single projective measurement along a mode which partially overlaps with all the input modes. Over the last decades, tremendous progress has been performed in terms of generation, manipulation and distribution of quantum states, thanks to the possibility to rely on photons as quantum information carriers. These schemes usually rely on individual conditioning steps, which can be photon counting \cite{Eaton2019} or homodyne conditioning  \cite{Lund2004,Vasconcelos2010,Etesse2014,Etesse2015,Sychev17,Weigand2018,Oh2018}.
%%%%%%%%%%%%%%%%%%%%%%%%%%%%%%%%%%%%%%%

\begin{abstract}
Disposing of simple and efficient sources for photonic states with non-classical photon statistics is of paramount importance for implementing quantum computation and communication protocols. In this work, we propose an innovative approach that drastically simplifies the preparation of non-Gaussian states as compared to previous proposals, by taking advantage from the multiplexing capabilities offered by modern quantum photonics tools. Our proposal is inspired by iterative protocols, where multiple resources are combined one after the other for obtaining high-amplitude complex output states. Here, conversely, a large part of the protocol is performed in parallel, by using a single projective measurement along a mode which partially overlaps with all the input modes. We show that our protocol can be used to generate high-quality and high-amplitude Schr\"odinger cat states as well as more complex states such as error-correcting codes. Remarkably, our proposal can be implemented with experimentally available resources, highlighting its straightforward feasibility.
\end{abstract}

%%%%%%%%%%%%%%%%%%%%%%%%%%%%%%%%%%%%%%%%
%\pacs{03.65.Ud, 03.67.-a, 42.50.Dv}

\maketitle
%%%%%%%%%%%%%%%%%%%%%%%%%%%%%%%%%%%%%%%%

\section*{Introduction}

Efficient coding of quantum information relies on the capability to generate high-quality states of light with non-classical photon statistics, both in the continuous-~\cite{Lvovsky2020} and discrete-variable~\cite{Zhong2020} regimes. In particular, states with negative Wigner functions have shown to be crucial for implementing sophisticated protocols and information processing, such as error-correction~\cite{Gottesman2001} and one-way quantum computing~\cite{Zheng2021}. Different strategies for generating such states were reported, relying for instance on high-order nonlinearity~\cite{Rempe2019,Kosuke2022,FLY22}, qubit interaction~\cite{Hastrup2021} and linear optical interaction ~\cite{Lund2004,Etesse2014,Weigand2018,Eaton2019}. A class of protocols of particular interest is that of iterative schemes \cite{Fiurasek2005,Etesse14,Etesse2014,Lund2004,Weigand2018,Eaton2019}, where photonic `resource' states are combined together in an iterative fashion, allowing for growing step-by-step the size of the states. However, the experimentally implemented protocols are to date restricted to few operations, due to protocol scalability issues arising from the simultaneous need for multiple heralding signals and from the increase of experimental overhead~\cite{Etesse2015,Sychev17}. This ultimately limits the average photon number in the generated state.\\
In this article, we propose a novel scheme to overcome the current experimental limitations, by relying on a single shot operation with a spectrally multiplexed resource. Our proposal takes advantage of the multimode aspect of state-of-the-art experimentally investigated quantum photonic sources in order to greatly enhance the capabilities of iterative generation protocols. This improvement occurs at two levels: (i) the number of operations to implement is reduced to one, and, as a consequence, (ii) the success probability of the overall operation is strongly increased. To illustrate our proposal, we show that Schr\"odinger cat state (SCS) and GKP-codes \cite{Gottesman2001} can be generated in an efficient manner, proving the validity of the approach, and opening a new path in the domain of non-Gaussian photonic state preparation.\\
In Sec.~\ref{sec_prot} we recall the principle of iterative protocols and propose a fully parallelized counterpart relying on spectral multimode quantum optics. Section \ref{sec_pargen} is devoted to the investigation of parallelized generation of cat states and GKP codes. Finally, in Sec.~III we show that our protocol can be used with experimental readily accessible ressources such as single photon Fock states.

\section{Multimode iterative protocol}
\label{sec_prot}
\subsection{Framework: iterative protocols}
\begin{figure}[!h]
\begin{center}
\includegraphics[width=6cm]{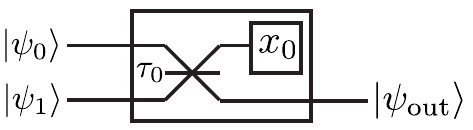}
\caption{Elementary brick of an iterative protocol.}
\label{protelem}
\end{center}
\end{figure}

The protocol proposed in this paper takes its essence from iterative generation protocols, an elementary brick of which is represented in Fig. \ref{protelem}.\\

\begin{figure}[!h]
\begin{center}
\includegraphics[width=\columnwidth]{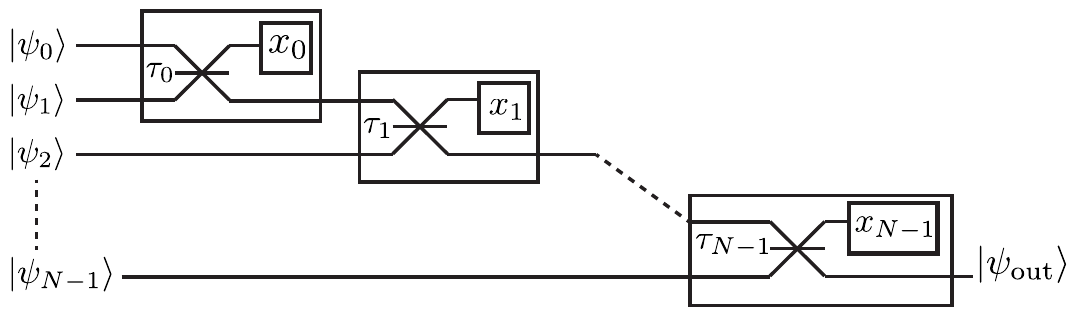}
\caption{A cascaded architecture for an iterative protocol.}
\label{protcasc}
\end{center}
\end{figure}

Such operation consists in mixing two input `resource' states, $\ket{\psi_0}$ and $\ket{\psi_1}$, on a beamsplitter with transmittivity $\tau_0$, to herald the generation of an ouput state $\ket{\psi_{\rm out}}$. The heralding process can be performed in different ways~\cite{Etesse2014,Eaton2019}, and even if the results derived here can easily be extended to the photon counting case, we focus in the following on homodyne conditioning. In this scenario, one of the two output arms of the beamsplitter is measured by a homodyne detector (along a certain quadrature $x_\theta$). When the measurement outcome is close enough to an assigned value ($x_\theta=x_0$ on Fig.\ref{protelem}), the target output state is successfully heralded. The strength of such a protocol is that the average photon number of the output state can be higher than the one of the individual input states. Accordingly, it can be used to `breed' complex photonic states by cascading the protocol using $\ket{\psi_{\rm out}}$ as one of the input of a subsequent operation.\\
Figure \ref{protcasc} shows such a configuration, where states of increasing size are combined with resource states; tree-like configurations are also commonly investigated, thanks to the parallelization capabilities they offer \cite{Etesse14}.
This idea is at the core of the `cat breeding' scheme, proposed in \cite{Lund2004}, where balanced beamsplitter ($\tau=1/\sqrt{2}$) and $x_i=0$ homodyne conditioning are used \cite{Takeoka2007,Brask2010}. In this case, mixing two SCSs allows to create a single SCS with an amplitude increased by a factor $\sqrt{2}$. Iterating this protocol $N$ times then allows to create a SCS with an amplitude amplified by a factor $2^{N/2}$. An alternative manner has also been proposed, where the input cat states were replaced by single photons, allowing the production of squeezed SCSs, robuster to losses \cite{Etesse2014,LeJeannic2018}. By modifying the quadrature along which the homodyne conditioning is performed ($p=0$ homodyne conditionings), GKP codes can be generated all-optically \cite{Vasconcelos2010,Etesse14}. More generally, all superpositions of up to $N$ photons can be generated by properly adjusting the beamsplitter reflectivities and homodyne conditionings \cite{Etesse2014}.\\
Proof-of principle experimental validations are currently limited to single stage implementations due to scalability issues \cite{Etesse2015,Sychev17}: in bulk optics, increasing the number of stages by increasing the number of input state sources indeed leads to intractable complexities. Temporal multiplexing represents a promising alternative that is also investigated~\cite{Bouillard2019}, but could ultimately lead to a decrease in the overall generation rate of the states. Integrated optics, on the other hand, offers an attractive solution to address spatial scalability issues, but inevitable losses associated with such platforms should be carefully taken into consideration \cite{Alibart2016}. \\
We propose in this article to rely instead on spectral multiplexing, which offers an increased generation efficiency as well as noticeable experimental simplifications. 

%In the next paragraph, we recall a few important notions linked with spectral multimode quantum optics.

\subsection{Multiplexed single-stage generation protocol}
\begin{figure}[!h]
\begin{center}
\includegraphics[width=\columnwidth]{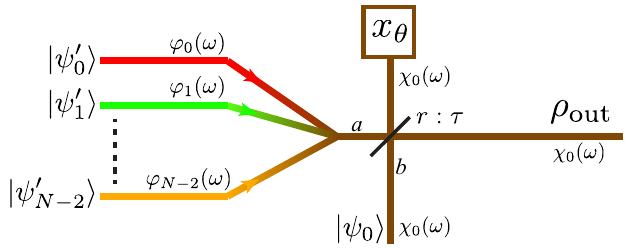}
\caption{The multiplexed generation protocol (see text for description).}
\label{protmult}
\end{center}
\end{figure}

The main idea of the proposal, sketched in Fig.\ref{protmult}, consists in the following key ingredients. To begin with, a `seed' state $\ket{\psi_0}$ initially populates a spectral mode $\chi_0(\omega)$ in which the quantum engineering process is to be performed. In output of the protocol, we will focus on the state $\rho_{\rm out}$ that populates this same mode $\chi_0(\omega)$. Subsequently, we select a collection of states $|\psi'_0\rangle$, $|\psi'_1\rangle$, ... $|\psi'_{N-2}\rangle$ respectively in modes $\varphi_0(\omega)$, $\varphi_1(\omega)$, ... $\varphi_{N-2}(\omega)$ that are used to `feed' the protocol, so as to grow the overall size of the output state $\rho_{\rm out}$.  As will be discussed with a concrete example, the spectral modes $\varphi_i(\omega)$ have to overlap $\chi_0(\omega)$ in a clever way, so that each $\ket{\psi'_i}$ contributes in an efficient manner to the generation protocol. 
This collection of states is merged in a single spatial mode and sent to the first input port (port $a$) of a beamsplitter of transmittivity $\tau$ so as to interfere with the seed state $\ket{\psi_0}$ sent to the other input port of this beamsplitter (port $b$). Importantly, a perfect spatial mode matching of the two states populating modes $a$ and $b$ must be guaranteed. After interaction, one output port of the beamsplitter is measured by a homodyne detector along mode $\chi_0(\omega)$, to perform the desired conditioning $x_\theta=x_0$, for successfully heralding the generation of a state $\rho_{\rm out}$ on the other output port.

A major advantage of this method is that it allows mixing $N$ states in a single heralding step, therefore drastically increasing the success probability of the protocol and reducing the experimental complexity as compared with the standard spatial multiplexing strategy, as will be quantitatively shown later.\\ 
The multimode description that we use complies with realistic implementations of gaussian state sources, such as, for instance, multimode squeezed vacuum sources \cite{Wasilewski06}. It is now indeed well established that spontaneous parametric down-conversion (SPDC) through three-wave mixing process produces a collection of squeezed vacuum states in separate orthogonal spectral modes  $\{\varphi_k(\omega)\}_{k\in\mathbb N}$, each with a different squeezing parameter \cite{Branczyk2010,RomanRodriguez2021}.\\
More generally, each spectral mode $\varphi_k(\omega)$ can be associated with a bosonic anihilation operator $\hat{A}_k$ :
\begin{align}
\hat{A}_k=\int\varphi_k^*(\omega)\hat{a}(\omega)d\omega,
\end{align}
$\{a(\omega)\}$ being the bosonic operators of individual spectral components  at the SPDC output.
When two different sets of modes $\{\varphi_k(\omega)\}_{k\in\mathbb N}$ and $\{\chi_l(\omega)\}_{l\in\mathbb N}$ are involved, one can express their `likeness' by using the overlap formula
\begin{align}
c_{k,l}=\int\varphi_k^*(\omega)\chi_l(\omega)d\omega.
\end{align}
This formula allows switching from one mode basis to the other thanks to
\begin{align}
\varphi_k(\omega)=\sum_lc^*_{k,l}\chi_l(\omega).
\label{eq_baschang}
\end{align}
We show in the next paragraph that the proposed protocol can be used to generate highly-fidel SCSs and GKP codes with increased success probabilities compared to standard iterative protocols.

\section{Paralellized generation of complex photonic states}
\label{sec_pargen}
\subsection{Parallel cat `breeding'}
\label{subsec_catbreed}
\begin{figure*}
\begin{center}
\subfigure[]{
\includegraphics[width=0.49\columnwidth]{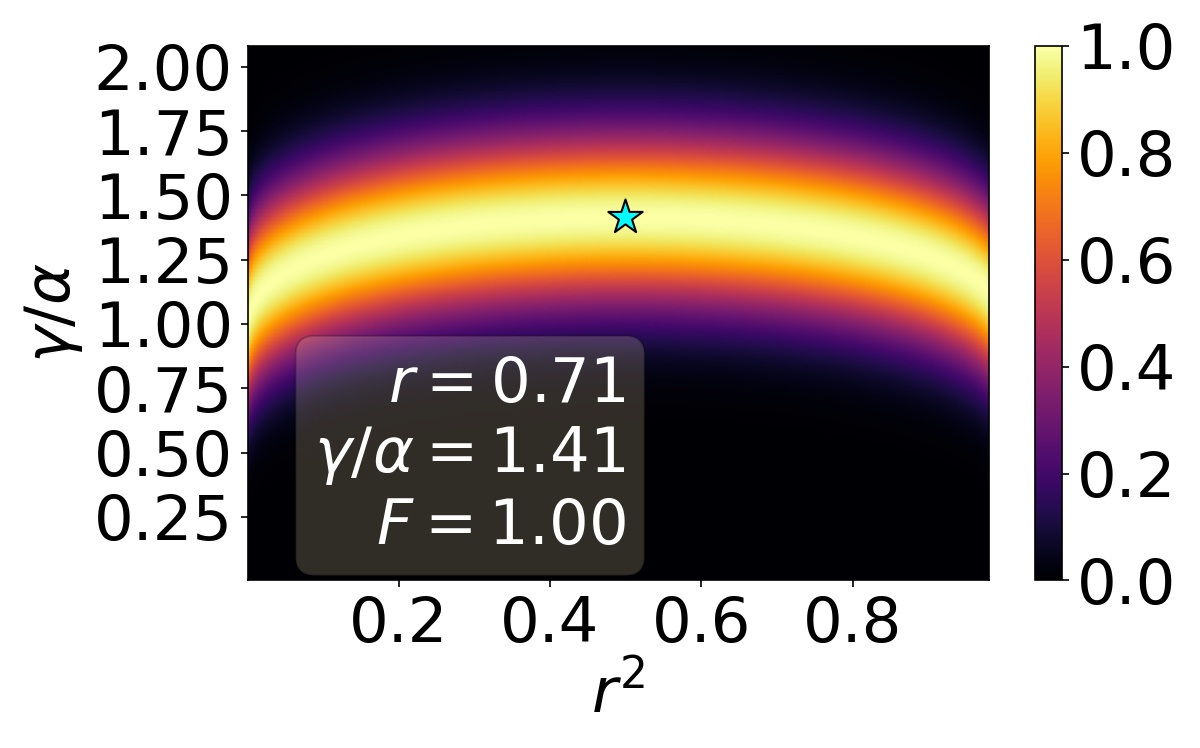}}
\subfigure[]{
\includegraphics[width=0.49\columnwidth]{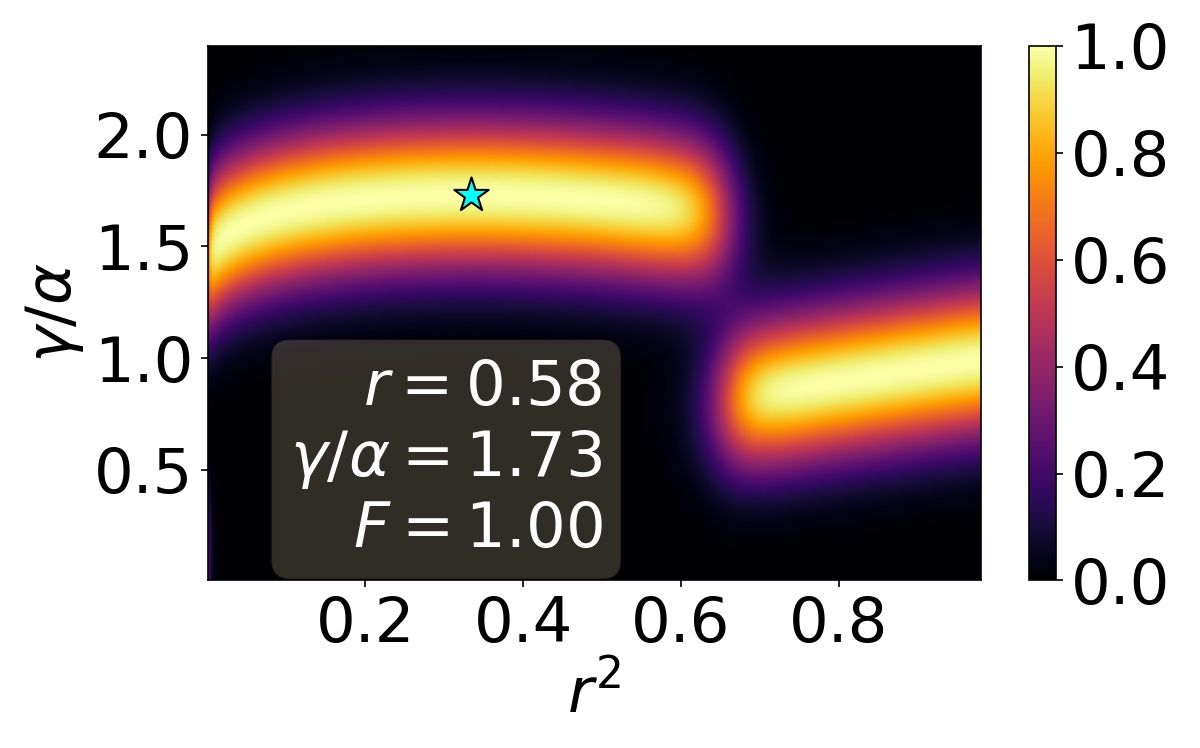}}
\subfigure[]{
\includegraphics[width=0.49\columnwidth]{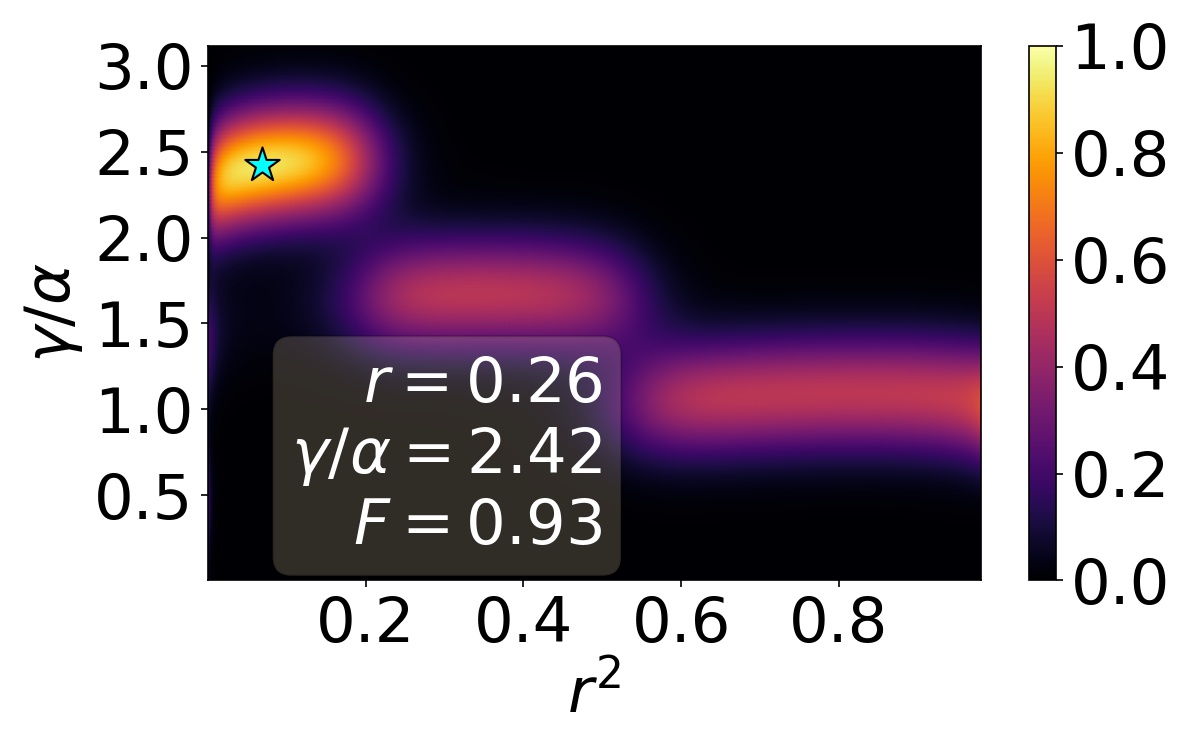}}
\subfigure[]{
\includegraphics[width=0.49\columnwidth]{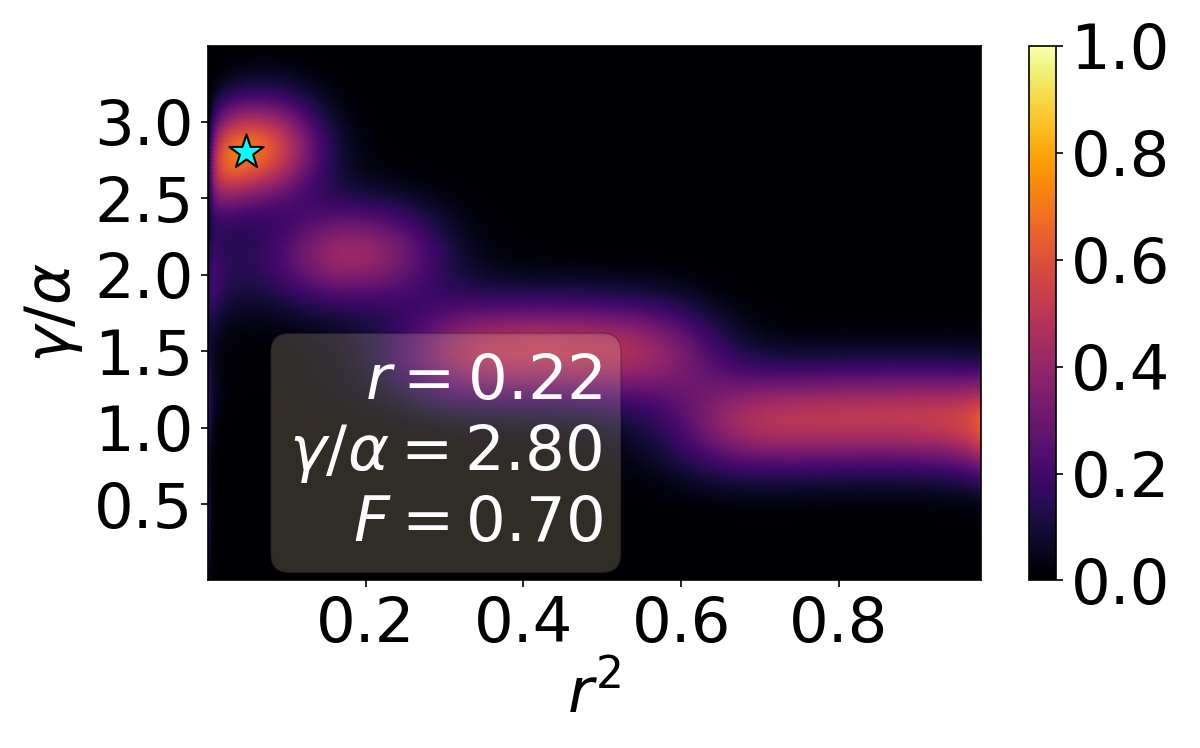}}
\subfigure[$N = 2$]{
\includegraphics[width=0.49\columnwidth]{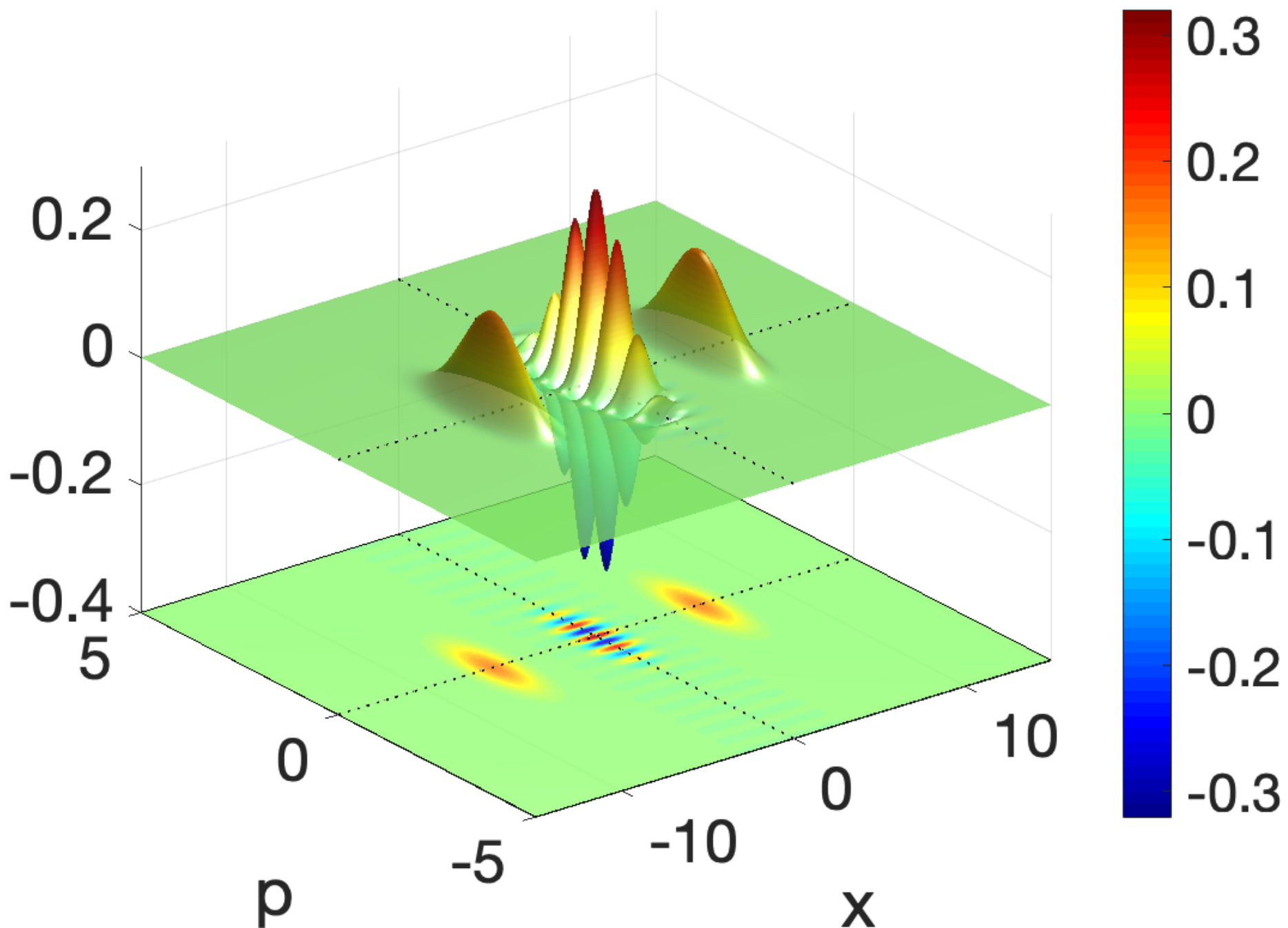}}
\subfigure[$N = 3$]{
\includegraphics[width=0.49\columnwidth]{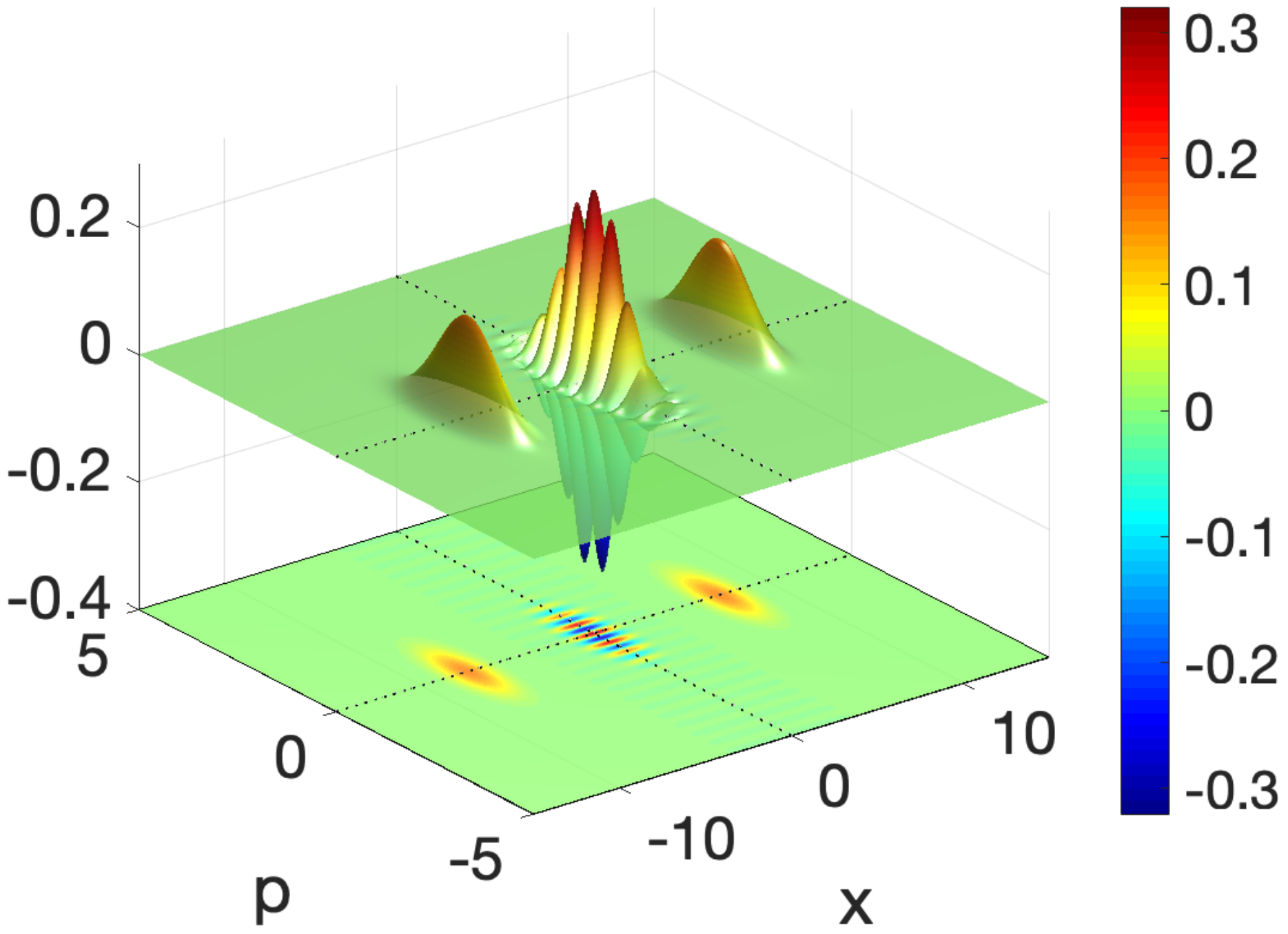}}
\subfigure[$N = 6$]{
\includegraphics[width=0.49\columnwidth]{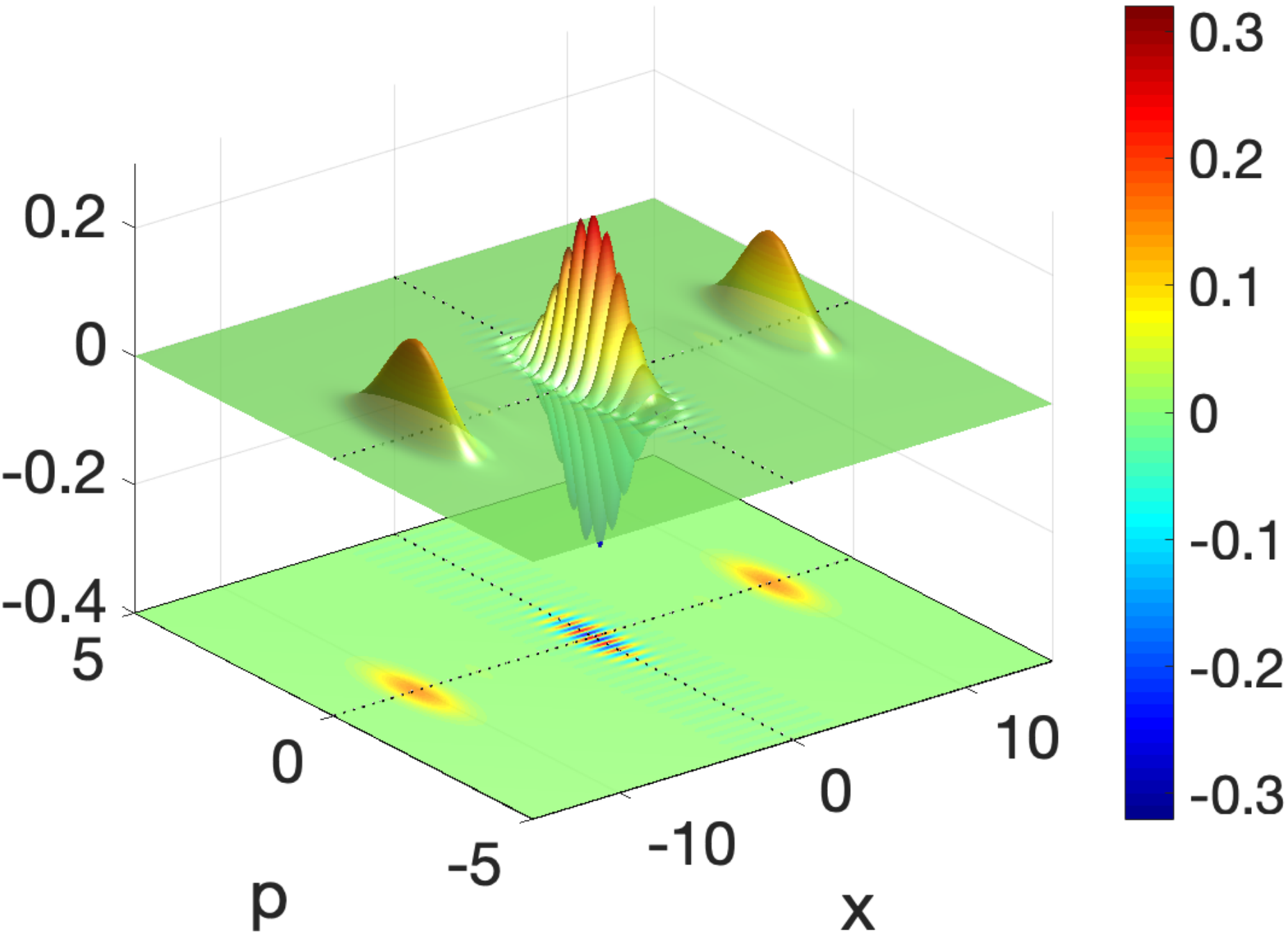}}
\subfigure[$N = 8$]{
\includegraphics[width=0.49\columnwidth]{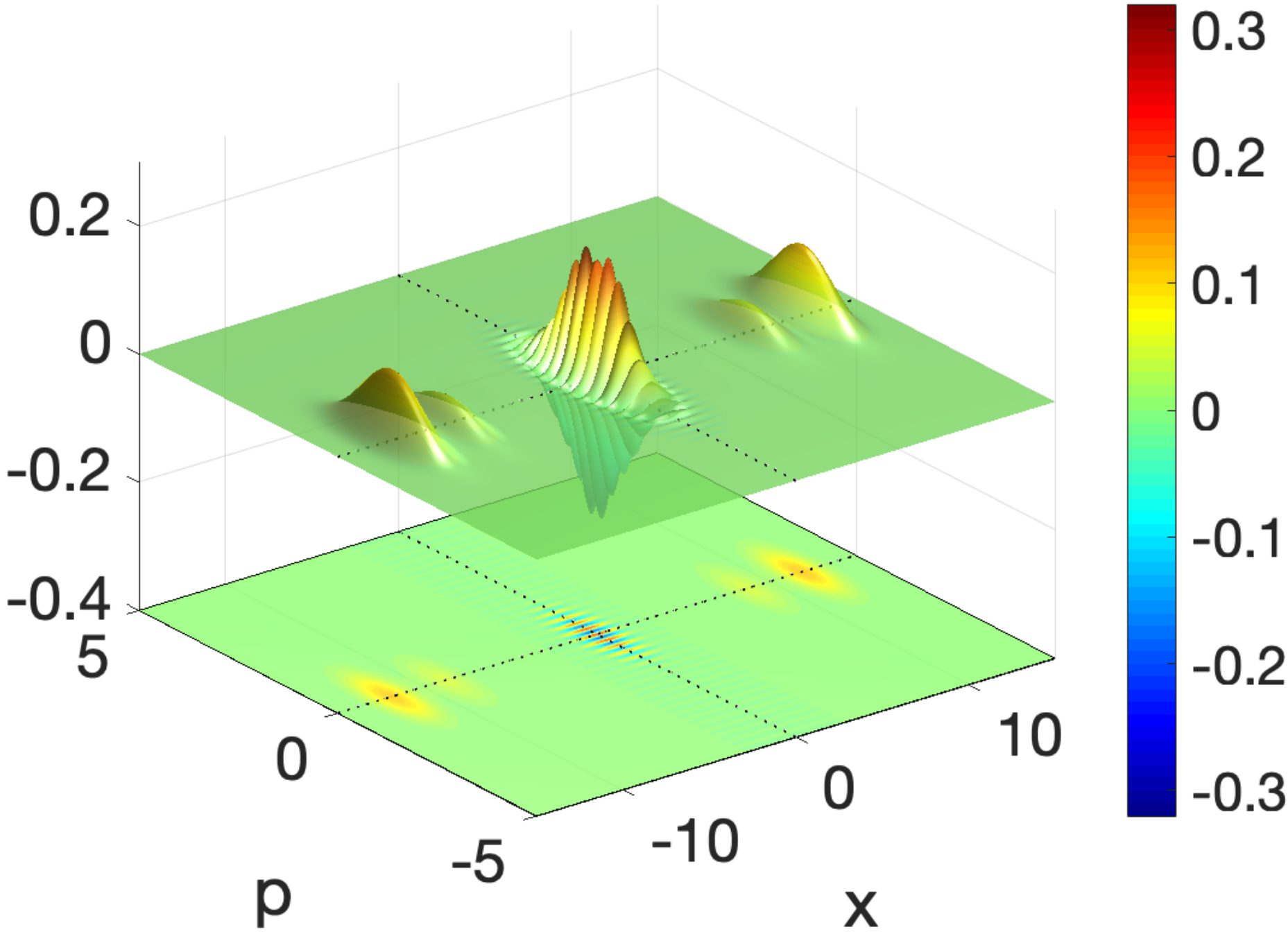}}
\caption{Fidelity of the output state with a cat state of amplitude $\gamma$ as a function of the beamsplitter reflectivity $r$, for (a) $N=2$, (b) $N=3$, (c) $N=6$ and (d) $N=8$ input cat states of amplitude $\alpha=3$. In each plot, a star spots the highest fidelity point, and the corresponding parameters are given in the insets. The Wigner functions corresponding to this maximal fidelity point are represented in (e), (f), (g), (h) respectively.}
\label{cat_reflect}
\end{center}
\end{figure*}
The previously described scheme can be applied to the generation of large amplitude SCSs, inspired by cat `breeding' iterative protocols. To perform such a generation, we simply consider here $N\geqslant2$ identical SCS input states of amplitude $\alpha$ in the $N-1$ input modes $\varphi_k(\omega)$:
\begin{align}
\forall k,\in[\![0,N-2]\!],\ \ket{\psi_k'}=\ket{\psi_0}=\frac{1}{\mathcal{N}}\left(\ket{\alpha}+\ket{-\alpha}\right),
\end{align}
where $\mathcal{N}=\sqrt{2\left(1+e^{-2|\alpha|^2}\right)}$ is a normalization factor. Note that we restrict the discussion to even cat states, but similar reasoning is also valid with odd cat states, as treated in the Suppl. Mat.. Then we set that each mode $\varphi_k(\omega)$ is of equal contribution in the generation process, namely $\forall k\in[\![0,N-2]\!]$:
\begin{align}
c_{k,0}=\int\varphi_k^*(\omega)\chi_0(\omega)d\omega=\frac{1}{\sqrt{N-1}}.
\label{part_overlap}
\end{align}
In this way, each input `feeding' cat $\ket{\psi_k'}$ will `breed' the input seed cat $\ket{\psi_0}$ by the same amount.
Subsequently, homodyne conditioning is performed along the $x$ quadrature (in-phase with the SCS) around $x=0$, within a given acceptance window $[-\Delta x/2,\Delta x/2]$. An analytic formula for the output Wigner function (see Suppl. Mat.) can then be derived, allowing us to predict the performances of the protocol as a function of key parameters such as $r$, $N$, $\alpha$ and the heralding width $\Delta x$. Many different metrics can be used to quantify the distance between an output state and a target state. Here, the fidelity of $\rho_{\rm out}$ with the nearest ideal SCS is used (amplitude $\gamma$).
\subsubsection*{Beamsplitter reflectivity}
The aforementioned fidelity of the output state is plotted in Figs. \ref{cat_reflect}(a) to (d) for $N=$2, 3, 6 and 8  input SCSs of amplitude $\alpha=3$, as a function of the beamsplitter reflectivity $\tau$ and of the output state amplitude $\gamma$. These plots clearly reveal that the output amplitude of the cat state in mode $\chi_0(\omega)$ that corresponds to optimal fidelities with $\rho_{\rm out}$ grows with the number of input states. This successfully proves the breeding of the input SCSs. Figure~\ref{cat_reflect} also shows that the reflectivity should be optimized in a non-trivial way, according to the number of input breeding states, in order to maximize the output state fidelity with high-amplitude SCSs. 

To gain a qualitative intuition on the quality of the multimode breeding operation, the Wigner functions of the corresponding output states are plotted in Figs.\ref{cat_reflect} (e) to (h), showing their strong non-classical signature, as the Wigner functions contain oscillations with negative values. To be more quantitative on this result, the numerical values of fidelity, optimal reflectivity and output SCS amplitudes are also written on Figs. \ref{cat_reflect} (a) to (d), revealing the very high fidelities of the output states with high-amplitude SCSs. In addition, it confirms that the amplitude of the output SCS increases with $N$, with a scaling of the order of $\gamma~\sim~\sqrt{N}\alpha$.
\subsubsection*{Amplitude of the input states}
\begin{figure}[!h]
\begin{center}
\includegraphics[width=0.9\columnwidth]{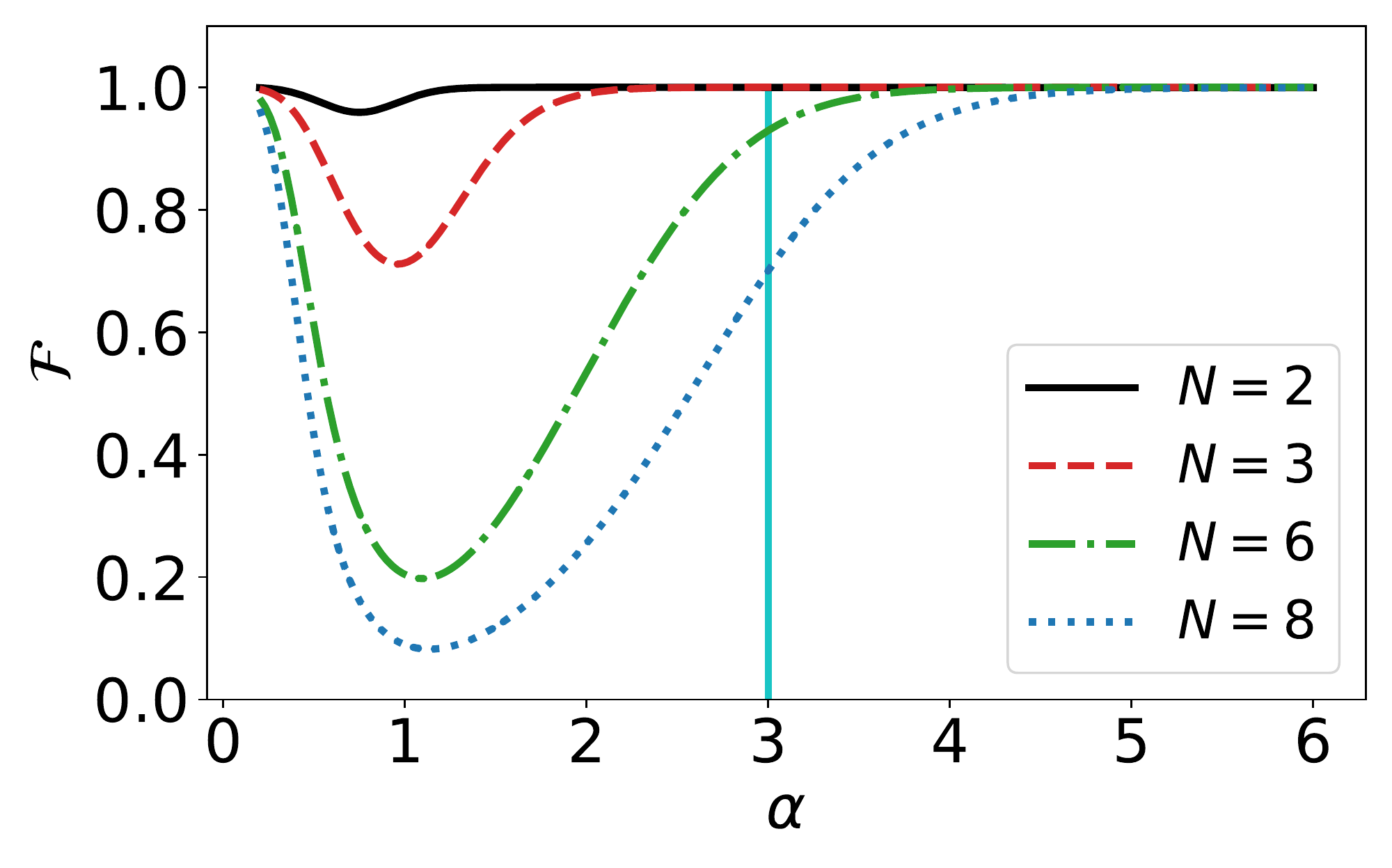}
\caption{Fidelity of the output state with the nearest SCS as a function of $\alpha$, for different $N$. The cases represented in fig.\ref{cat_reflect} correspond to the vertical blue line ($\alpha=3$).}
\label{dep_alpha}
\end{center}
\end{figure}

The protocol is also sensitive to the amplitude of the input cat states. This is illustrated in Fig.~\ref{dep_alpha}, where the output state fidelity with the nearest cat state is represented as a function of the input SCS amplitude $\alpha$, for different numbers of input states $N$. Two important trends shall be noticed: (i) for a given $\alpha$, increasing $N$ decreases the overall output state fidelity. However, (ii) increasing $\alpha$ for a given $N$ leads to an increased global fidelity. For instance, to create states with 99\% fidelity with $N=2$, an input amplitude $\alpha=1.13$ is sufficient, whereas $\alpha=2.49$ is required to reach the same fidelity with $N=4$ input states.
 In the case of even cat states that are considered here, there is also an increase of the fidelity for vanishing $\alpha$ simply because vacuum is a perfect even cat state with zero amplitude, a case of no interest for us.

\subsubsection*{Comparison with the iterative cat breeding protocol}
\begin{figure}[!h]
\begin{center}
\includegraphics[width=0.9\columnwidth]{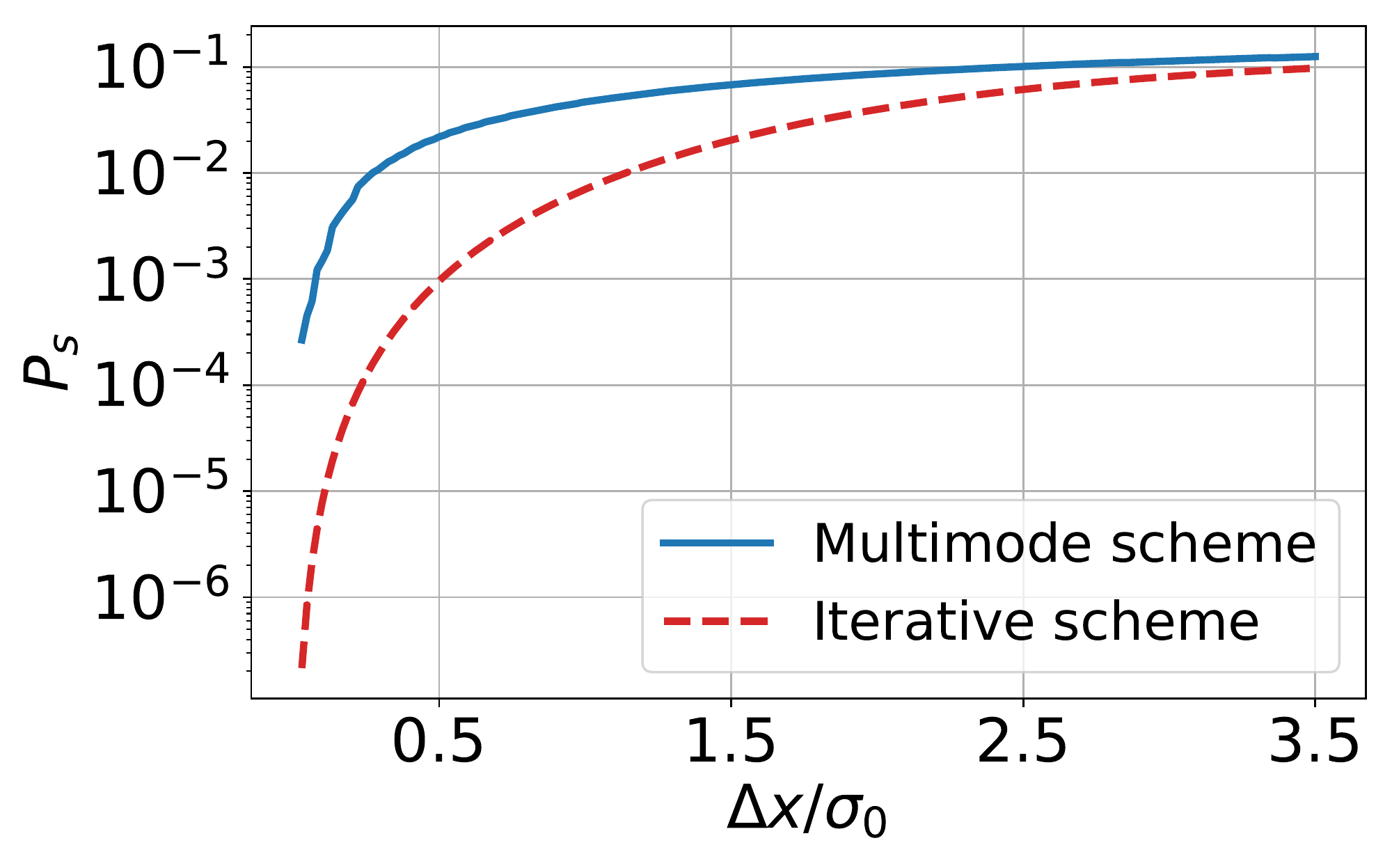}
\caption{Comparison between the success probability of the multimode scheme and the iterative scheme for $N=4$, as a function of the heralding window $\Delta x$ normalized to the shot noise standard deviation $\sigma_0$.}
\label{psucc_compare}
\end{center}
\end{figure}

Interestingly, we can draw a direct parallel with the iterative cat breeding protocol, where $N-1$ heralding steps are performed on $N$ input resources. The main difference, however, is that the multimode protocol that we propose here is inevitably accompanied with information loss due to partial overlap of the spectral modes $\varphi_k(\omega)$ with the output mode of interest $\chi_0(\omega)$. Equation (\ref{part_overlap}) directly shows that each mode $\varphi_i(\omega)$ overlaps only partially with $\chi_0(\omega)$, therefore all of the components of the input states $\ket{\psi'_i}$ along $\chi_{l>0}(\omega)$ are purely lost during the process, where $\{\chi_l(\omega)\}_{l\in\mathbb N}$ constitutes a complete spectral basis. Therefore, the output state $\rho_{\rm out}$ cannot be a pure state as long as more than two modes are populated in total.\\
To give an intuitive model, we propose a simple parallel between spatial and spectral modes: the operation depticted in Fig.~\ref{protmult} by the merging arrows from modes $\varphi_k(\omega)$ to $\chi_0(\omega)$ can actually be simply modeled as a $N-1$ input / $N-1$ output beamsplitter transformation, where $N-2$ outputs are lost in the environment. This transformation is exactly the one depicted in Fig.~\ref{protcasc} for input states $\ket{\psi_0}$ to $\ket{\psi_{N-2}}$, with $\tau_i=1/\sqrt{i+2}$ so as to obtain $c_{0,k}=1/\sqrt{N-1}$. In other words, our protocol is equivalent to the iterative breeding scheme of Fig.~\ref{protcasc}, where only the last heralding step $x=x_{N-1}$ is performed and all the other ones are simply ignored. The direct consequence for this is that the overall success probability of the protocol is increased by several orders of magnitude as compared to the iterative scheme. This is illustrated in Fig.~\ref{psucc_compare}, where the success probability of our scheme for input SCS amplitudes of $\alpha=3$ and $N=4$ is plotted, in comparison with the iterative scheme of Fig.\ref{protcasc} in the case where all the heralding widths are identical. The cost obviously is the overall fidelity, which is degraded in the multimode case, but nevertheless remains very high, as we have seen in Fig.~\ref{cat_reflect}. The multimode protocol therefore takes advantage of the robustness of the breeding protocol for imperfect heralding at sufficiently high SCS amplitudes.
\subsection{GKP states preparation}
In addition to the generation of SCSs, our protocol can also advantageously be applied to for the generation of more complex states such as the GKP codes \cite{Gottesman2001}. These states were first introduced as a means to encode information in an oscillator, with an error-correcting capability. They were recently generated and used in trapped ions~\cite{Fluhmann2019} and superconducting circuits~\cite{CampagneIbarcq2020}, but still remain out of reach of all-optical protocols. Here, we can use the same protocol as the one depicted in Sec. \ref{subsec_catbreed} ($N-1$ input cat states in modes $\varphi_k$ and one cat state in mode $\chi_0$, all of amplitude $\alpha$), with the only difference that homodyne conditioning is performed along $x_{\pi/2}:=p=0$. Such conditioning has already been investigated in the context of all-optical generation of GKP with iterative protocol~\cite{Vasconcelos2010,Etesse14}. We show here that it can be transposed to our multiplexed approach. As a reminder, the wavefunctions of approximate GKP codes $\ket{\tilde{0}}$ and $\ket{\tilde{1}}$ take the form
\begin{subequations}
\begin{align}
\langle x \vert \tilde{0} \rangle& = \dfrac{1}{\mathcal{N}_0} \sum_{k \in \mathbb{Z}} e^{-\frac{s_1^2}{2} (ka)^2} e^{-\frac{1}{2 s_2^2}(x - k a)^2}\\
\langle x \vert \tilde{1} \rangle &= \dfrac{1}{\mathcal{N}_0} \sum_{k \in \mathbb{Z}} e^{-\frac{s_1^2}{2} (k + 1 / 2)^2 a^2} e^{-\frac{1}{2 s_2^2}\left(x - (k + 1 / 2) a \right)^2},
\end{align}
\end{subequations}
where $1/s_1$ gives the overall envelope size of the state, $s_2$ is the standard deviation of each tooth in the distribution, and $a$ is the distance between consecutive teeth along the $x$ quadrature.
\begin{figure}[!h]
\begin{center}
%\subfigure[]{
%\includegraphics[width=0.8\columnwidth]{Images/Wigner_GKP_K=3_alpha=3}}
\subfigure[]{
\includegraphics[width=0.8\columnwidth]{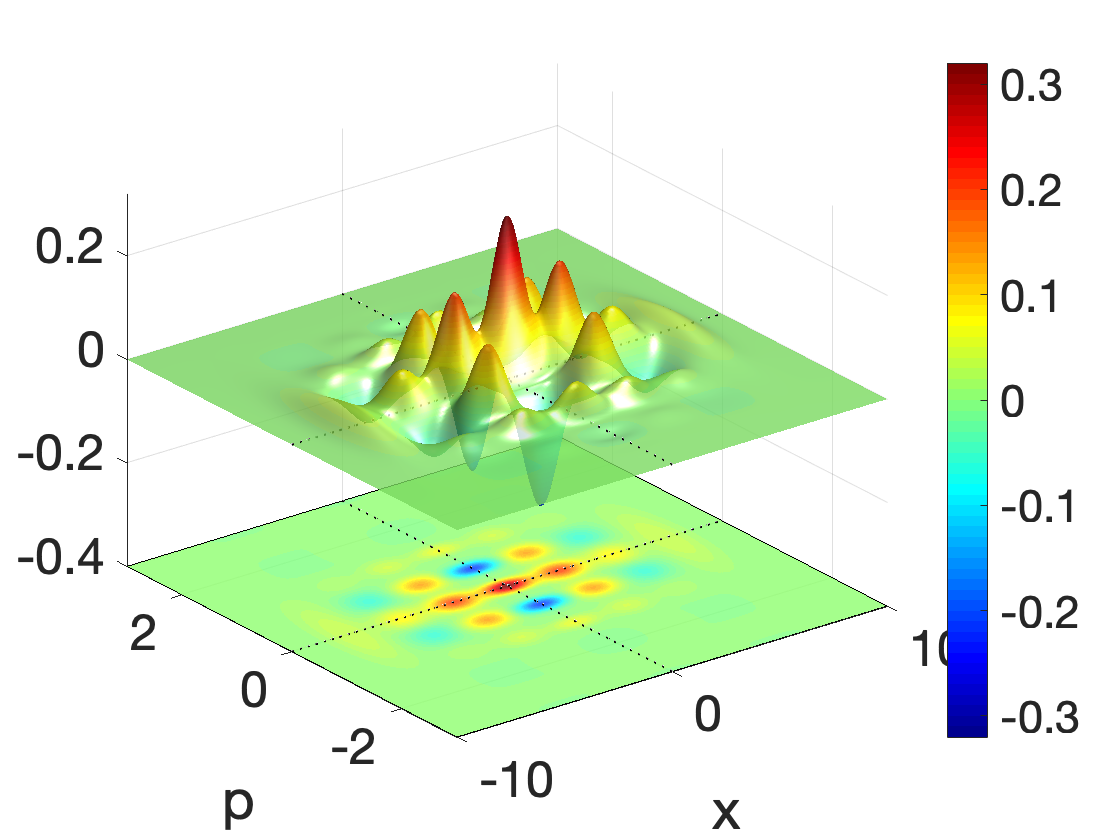}}
\subfigure[]{
\includegraphics[width=0.8\columnwidth]{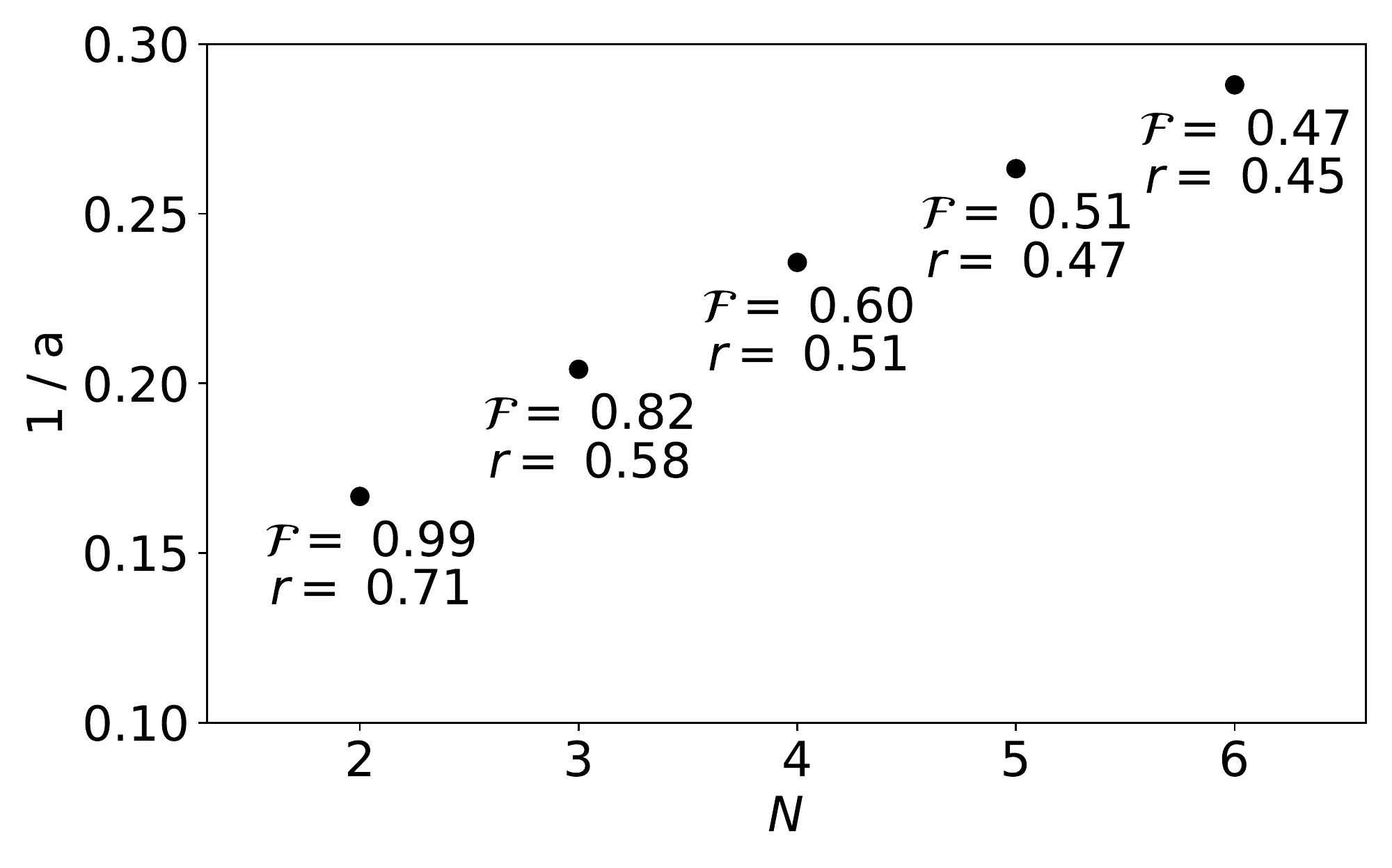}}
\caption{Performance of the multiplexed protocol for the generation of approximate GKP codes. (a) Wigner function of the output state with $N=3$ input SCSs of amplitude $\alpha=3$. (b) GKP state amplitude when the amount of input cat states $N$ is augmented.}
\label{GKP_perfs}
\end{center}
\end{figure}
Figure \ref{GKP_perfs}(a) shows the shape of the Wigner function of the output state for $N=3$, revealing the grid structure specific to GKP codes. Such likeness is, here again, quantified by the fidelity with the nearest GKP code, that we plot in Fig. \ref{GKP_perfs}(b). This figure shows that augmenting the amount of input resources indeed increases the size of the output state along the opposite quadrature ($1/a$), while requiring an optimization of $r$. Similarly to the SCS case, it also shows that the fidelity decreases with $N$.
%\begin{figure}[!h]
%\begin{center}
%\includegraphics[width=0.8\columnwidth]{Images/Fidelity_GKP_alpha=3}
%\caption{\je{Plutot tracer $1/a$ en fonction de $K$, qui donne que l'écartement des pics selon $p$ devient de plus en plus grand}}
%\label{GKP_ampli}
%\end{center}
%\end{figure}

%
%Un rappel du schema historique puis parallelise\\
%Performances : proba de succes, strategie de conditionnement avec reflectivite optimale

%\begin{figure*}[!h]
%\begin{center}
%\subfigure[]{
%\includegraphics[width=8cm]{Images/alpha3.png}}
%\subfigure[]{
%\includegraphics[width=8cm]{Images/alpha5.png}}
%\caption{}
%\label{}
%\end{center}
%\end{figure*}

\section{Proposal with experimentally accessible resources}
%\label{sec_exp}
\begin{figure*}
\begin{center}
\subfigure[]{
\includegraphics[width=0.45\columnwidth]{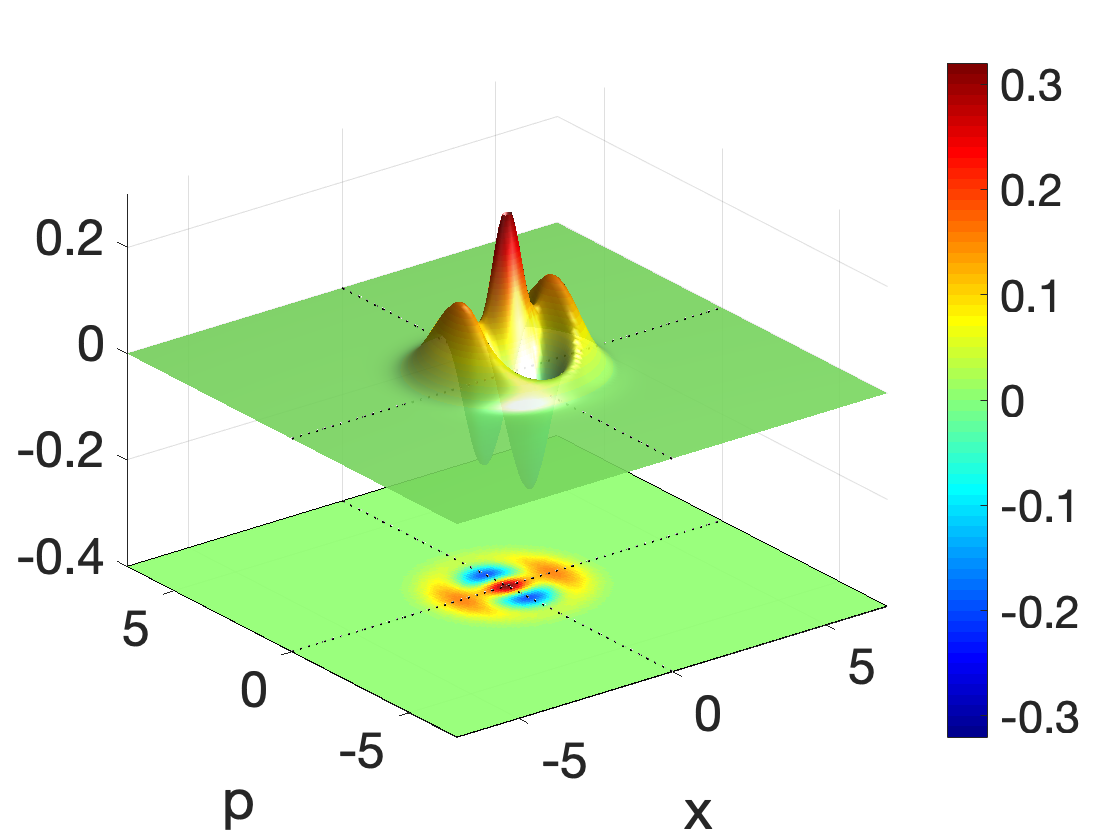}}
\subfigure[]{
\includegraphics[width=0.45\columnwidth]{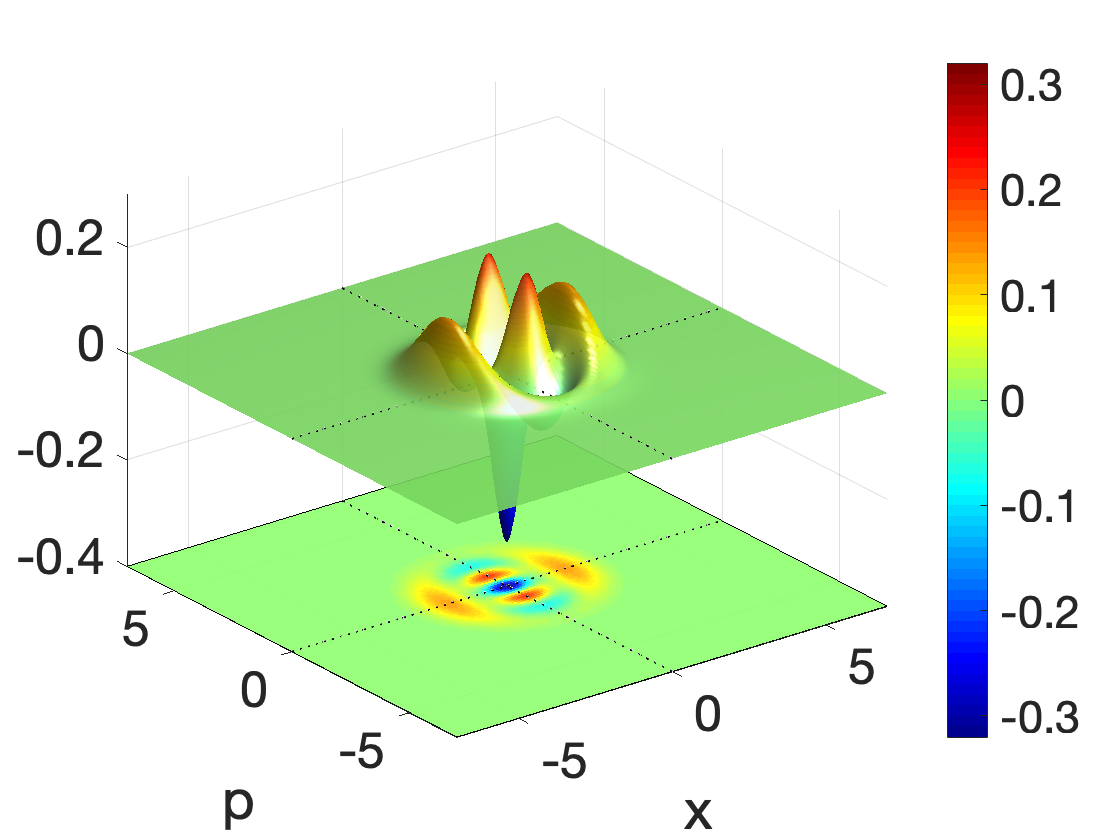}}
\subfigure[]{
\includegraphics[width=0.45\columnwidth]{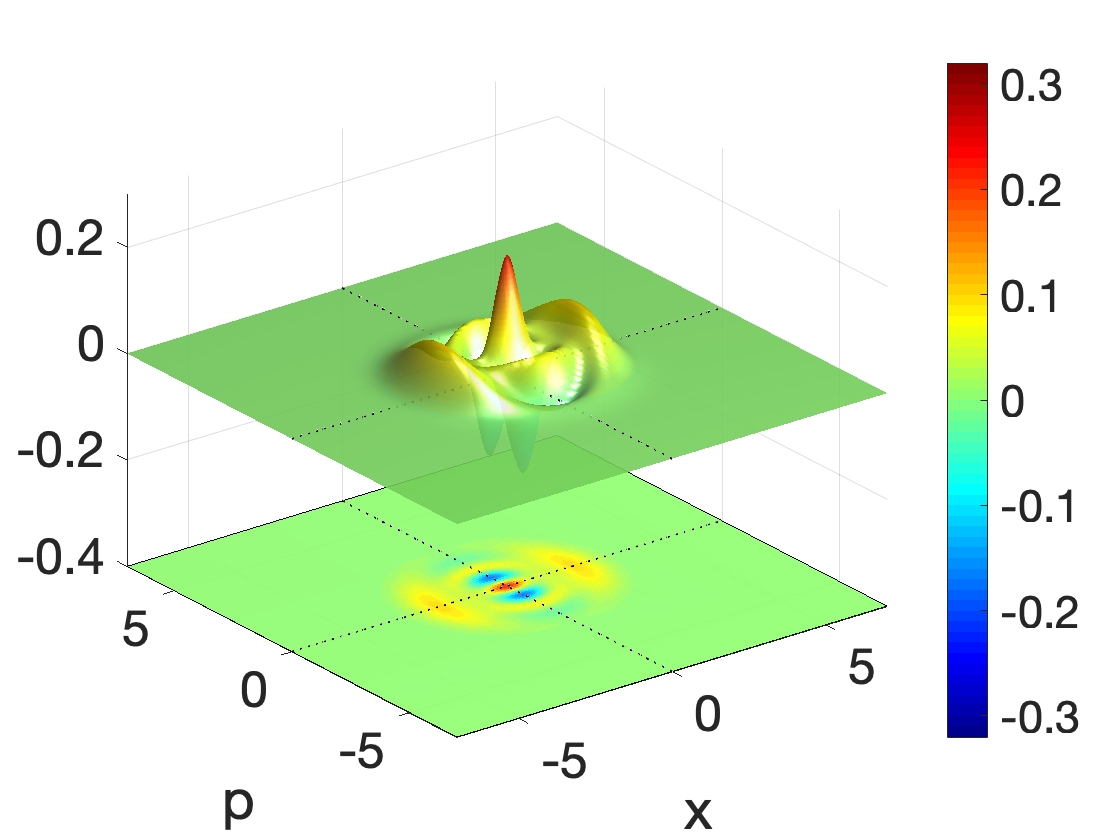}}
\subfigure[]{
\includegraphics[width=0.45\columnwidth]{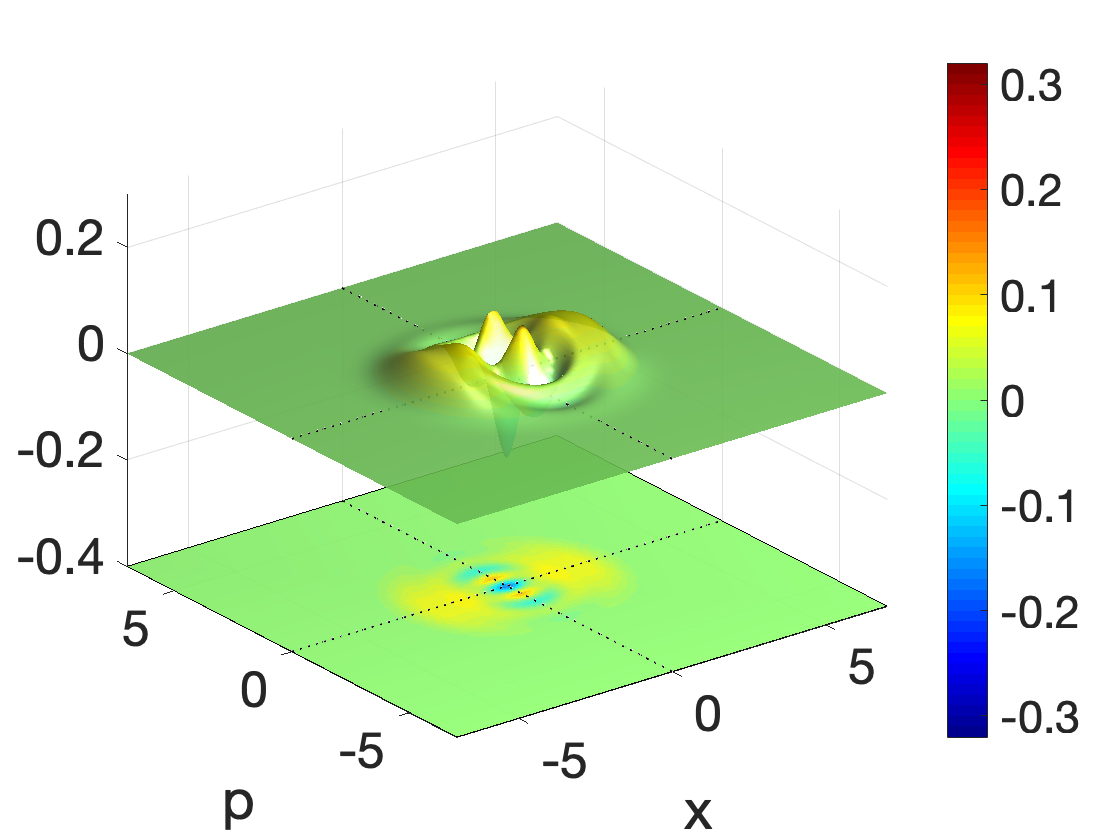}}

\caption{Wigner functions of the output state of the protocol fed by single photon Fock states, with (a) $N=2$, (b) $N=3$, (c) $N=4$ and (d) $N=5$.}
\label{sing_phot}
\end{center}
\end{figure*}
An important aspect of our strategy is that it can directly be implemented with already accessible experimental resources. Even if high-amplitude SCSs are not trivial to generate experimentally, good approximation of them at low amplitudes (also called Schr\"odinger `kitten' states) are commonly produced states. Moreover, the breeding principle has been experimentally demonstrated with kittens, showing the validity of the approach \cite{Sychev17}. In addition, single photon Fock states also constitute a valid alternative to them, as theoretically and experimentally demonstrated \cite{Etesse14,Etesse2015}, to generate squeezed SCS-like output states with high fidelity. It is worth noting that squeezing the output states constitutes a high advantage, as it can increase their robustness to losses~\cite{Ourjoumtsev2006,LeJeannic2018}. Accordingly, the results presented here will focus on input single photon Fock states. Very similar results are expected with kitten states.\\
We recall that our protocol requires the use of a multiplexed input state source. This can simply be achieved by taking advantage of the spectral tailoring possibility that processes like spontaneous parametric down-conversion offer~\cite{Branczyk2010}. Indeed, depending on the pumping and phase-matching conditions, two-mode squeezing can be generated along a large number of spectral modes, offering the ideal platform for multiplexed generation~\cite{RomanRodriguez2021}. Then, single photon Fock states are simply heralded by performing a projective measurement over one of the two mode squeezing subsystems. This method, although very commonly used for the generation of high-quality single photon Fock states for a couple of decades now, has recently been extended experimentally to high-dimension mode-selective quantum state tailoring~\cite{Ra2020}. This allows to tweak the supermode in which the single photon detection occurs, therefore heralding a single photon Fock state in a given supermode. Here we can take advantage of such capabilities, by using multiplexed photon sources to feed the protocol and generate complex photonic states in a single shot.\\
The output Wigner functions of the protocol fed by an increasing number $N$ of single photon Fock states are shown in Fig.~\ref{sing_phot}. They clearly reveal that the output state contains more and more oscillations at its center with increasing $N$. To be more quantitative, Tab. \ref{table_perf_SP} gathers the parameters of the closest SCS, where $\alpha$ is the amplitude, $s$ its compression parameter and $\mathcal{F}$ the fidelity between the two states. This table reveals that the amplitude of the state increases with $N$ for low values, but degrades after $N=4$. Further simulations show that this is true for larger $N$ as well. Such an evolution can qualitatively be explained by the phase invariance of the two states impinging the beamsplitter: as the only rotation symmetry breaking operation is the homodyne measurement, the fidelity of the created state with a highly non-phase invariant state decreases with its size.
\begin{table}[h!]
\begin{tabular}{|m{1cm}||m{1cm}|m{1cm}|m{1cm}|m{1cm}|}
\hline
&$N=2$&$N=3$&$N=4$&$N=5$\\
\hline
\hline
$\alpha$&1.63&2.03&2.30&2.12\\
\hline
$s$&0.66&0.65&0.66&0.74\\
\hline
$\mathcal{F}$&0.99&0.90&0.66&0.49\\
\hline
\end{tabular}
\caption{Fidelity of the output state with a squeezed SCS of amplitude $\alpha$ and squeezing $s$.}
\label{table_perf_SP}
\end{table}
It should also be pointed that the optimal generation condition (highest amplitude and fidelity) is fulfilled with the use of a low-reflectivity beamsplitter (see Suppl. Mat. for numerical simulations).\\
This analysis reveals that few-mode single photon states are relevant resources for our protocol in order to produce high-quality SCSs, essential brick in continuous-variable quantum information processing.

\section*{conclusion} We have proposed and discussed an innovative protocol for the generation of non-Gaussian states of light. Our scheme is inspired by iterative protocols but takes advantage of multiplexing capabilities of photonics resources. It relies on the use of multiple states in a single spatial mode but multiplexed in an other degree of freedom, mixed with a resource state whose mode partially overlaps all input states. By performing a homodyne conditioning after mixing the states at a beamsplitter, complex output states can be produced. The major strength of our proposal is that no cascade manipulation steps are required for the state breeding, therefore strongly simplifying the experimental implementation in a one-shot scheme, and allowing to access high-success rates. We have shown that frequency multiplexing can be used for the generation of high-quality and high-amplitude Schr\"odinger cat and GKP states, by using as input states small-amplitude Schr\"odinger cat states, or alternatively, readily accessible experimental resources such as multiplexed single photon Fock states. Note that current state-of-the-art non-linear sources offer the capability for generating such states by providing the native generation of frequency multiplexed single- and two-mode squeezing. Due to its high efficiency and simplified implementation as well as to the possibility of implementing it with already available experimental resources, we believe that our approach to non-Gaussian state preparation represents a major breakthrough in the domain of non-classical photonic state generation, and will open a wide range of applications.
\section*{acknowledgments}
This work has been conducted within the framework of the project HyLight (No. ANR-17-CE30-0006-01) granted by the Agence Nationale de la Recherche (ANR).

\bibliography{RefPD.bib}

\end{document}